\documentclass[reprint,twocolumn,superscriptaddress,prl,showpacs,amsmath,amssymb,aps]{revtex4-1}

\usepackage{natbib}	
\usepackage{graphicx,color,rotating}
\usepackage[latin1]{inputenc}
\usepackage{textcomp}
\usepackage{dcolumn}
\usepackage{bm}     
\usepackage{upgreek}
\usepackage{subdepth}  			
\usepackage{multirow}
\usepackage[normalem]{ulem}
\usepackage[latin1]{inputenc}

\usepackage{array}
\newcolumntype{L}[1]{>{\raggedright\let\newline\\\arraybackslash\hspace{0pt}}m{#1}}
\newcolumntype{C}[1]{>{\centering\let\newline\\\arraybackslash\hspace{0pt}}m{#1}}
\newcolumntype{R}[1]{>{\raggedleft\let\newline\\\arraybackslash\hspace{0pt}}m{#1}}

\usepackage{hyperref}						
\usepackage{xcolor}						
\hypersetup{							
    colorlinks,
    linkcolor={blue!80!black},
    citecolor={blue!80!black},
    urlcolor={blue!80!black}
}
\newcommand{\mr}[1]{\ensuremath{\mathrm{#1}}}
\newcommand{\myvec}[1]{\bm{#1}}
\newcommand{\ee}{\mathrm{e}}
\newcommand{\ii}{\mathrm{i}}

\newcommand{\avr}[1]{\big\langle #1 \big\rangle}
\newcommand{\abs}[1]{\big|{#1}\big|}

\DeclareMathOperator{\re}{Re}

\newcommand{\nablabf}{\boldsymbol{\nabla}}

\newcommand{\AAA}{\myvec{A}}

\newcommand{\BBB}{\myvec{B}}

\newcommand{\eee}{\myvec{e}}

\newcommand{\fff}{\myvec{f}}

\newcommand{\fffac}{\fff_\mathrm{ac}}

\newcommand{\III}{\myvec{I}}

\newcommand{\kc}{k_\mathrm{c}}
\newcommand{\kt}{k_\mathrm{t}}

\newcommand{\kcsqr}{k^{2_{}}_\mathrm{c}}

\newcommand{\nnn}{\myvec{n}}

\newcommand{\uuu}{\myvec{u}}

\newcommand{\VVV}{\myvec{V}}

\newcommand{\vvv}{\myvec{v}}

\newcommand{\calL}{\mathcal{L}}

\newcommand{\cp}{c_p}

\newcommand{\cv}{c_v}

\newcommand{\Dth}{D^\mathrm{th}}

\newcommand{\Eac}{E_\mathrm{ac}}
\newcommand{\Lac}{\calL_\mathrm{ac}}

\newcommand{\Pac}{P_\mathrm{ac}}

\newcommand{\kth}{k^\mathrm{th}}

\newcommand{\kthO}{k^\mathrm{th}_0}
\newcommand{\kthOsl}{k^\mathrm{th,sl}_0}
\newcommand{\kthOfl}{k^\mathrm{th,fl}_0}

\newcommand{\kthId}{k^{\mathrm{th},d}_1}

\newcommand{\kthIsl}{k^\mathrm{th,sl}_1}
\newcommand{\kthIfl}{k^\mathrm{th,fl}_1}

\newcommand{\kapT}{\kappa_T}

\newcommand{\kaps}{\kappa_s}

\newcommand{\alphap}{{\alpha_p}}
\newcommand{\alphapO}{{\alpha_{p0}}}
\newcommand{\alfP}{{\alpha_p}}

\newcommand{\delt}{\delta_\mathrm{t}}

\newcommand{\dels}{\delta_\mathrm{s}}

\newcommand{\etab}{\eta^\mathrm{b}}

\newcommand{\GamOcfl}{\Gamma^\mr{fl}_\mathrm{0c}}

\newcommand{\TOdfl}{T^{d,\mr{fl}}_0}
\newcommand{\TOdsl}{T^{d,\mr{sl}}_0}
\newcommand{\TO}{T_0}

\newcommand{\SIum}{\upmu\textrm{m}}

\newcommand{\SIMHz}{\textrm{MHz}}

\newcommand{\SIJ}{\textrm{J}}

\newcommand{\SIK}{\textrm{K}}

\newcommand{\SIm}{\textrm{m}}

\newcommand{\SImm}{\textrm{mm}}
\newcommand{\SImum}{\textrm{\textmu{}m}}

\newcommand{\SInm}{\textrm{nm}}

\newcommand{\SIs}{\textrm{s}}

\newcommand{\SIV}{\textrm{V}}

\newcommand{\nn}{\nonumber}

\newcommand{\eqlab}[1]{\label{eq:#1}}
\renewcommand{\eqref}[1]{Eq.~(\ref{eq:#1})}
\newcommand{\eqnoref}[1]{(\ref{eq:#1})}

\newcommand{\eqsnoref}[2]{(\ref{eq:#1}) and~(\ref{eq:#2})}

\newcommand{\figref}[1]{Fig.~\ref{fig:#1}}

\newcommand{\figlab}[1]{\label{fig:#1}}

\newcommand{\sigmabf}{\bm{\sigma}}

\newcommand{\taubf}{\bm{\tau}}

\newcommand{\cLsqr}{c^2_\mathrm{lo}}
\newcommand{\cTsqr}{c^2_\mathrm{tr}}

\newcommand{\fl}{\mathrm{fl}}

\renewcommand{\sl}{\mathrm{sl}}

\begin{document}

\title{A transition from boundary- to bulk-driven acoustic streaming due to nonlinear thermoviscous effects at high acoustic energy densities}
\author{Jonas Helboe Joergensen}
\email{jonashj@fysik.dtu.dk}
\affiliation{Department of Physics, Technical University of Denmark, DTU Physics Building 309, DK-2800 Kongens Lyngby, Denmark}

\author{Wei Qiu}
\email{wei.qiu@bme.lth.se}
\affiliation{Department of Biomedical Engineering, Lund University, Ole R\"{o}mers v\"{a}g 3, 22363, Lund, Sweden}

\author{Henrik Bruus}
\email{bruus@fysik.dtu.dk}
\affiliation{Department of Physics, Technical University of Denmark, DTU Physics Building 309, DK-2800 Kongens Lyngby, Denmark}

\date{21 December 2021}

\begin{abstract}
Acoustic streaming is studied in a rectangular microfluidic channel. It is demonstrated theoretically, numerically, and experimentally with good agreement, frictional heating can alter the streaming pattern qualitatively at high acoustic energy densities $\Eac$ above $500~\SIJ/\SIm^3$. The study shows, how as a function of increasing  $\Eac$ at fixed frequency, the traditional boundary-driven four streaming rolls created at a half-wave standing-wave resonance, transition into two large streaming rolls. This nonlinear transition occurs because friction heats up the fluid resulting in a temperature gradient, which spawns an acoustic body force in the bulk that drives thermoacoustic streaming.
\end{abstract}

\maketitle

Microscale acoustofluidic devices are used to manipulate and control microparticles and cells. In such devices, two main forces act on the suspended particles, the acoustic radiation force and the drag force due to acoustic streaming, which is a time-averaged flow caused by the inherent nonlinearities of fluid dynamics. Recent work has clarified many subtle details pertaining to the radiation force on mircoparticles, including thermoviscous effects \cite{Karlsen2015} and microstreaming \cite{Baasch2019}. Concurrently, similar progress has been made in the theory of acoustic streaming, especially regarding thermoviscous effects. The fundamental boundary-driven streaming caused by time-averaged forces in the oscillatory boundary-layer flow \cite{LordRayleigh1884}, and the fundamental bulk-driven streaming generated by the time-averaged dissipation of traveling waves \cite{Eckart1948}, have recently been supplemented by bulk-driven baroclinic \cite{Chini2014, Michel2021} and thermo\-acoustic \cite{Joergensen2021, Qiu2021} streaming, caused by an interplay between standing acoustic waves and steady temperature gradients. However, as noted in Refs.~\cite{Joergensen2021, Qiu2021}, the validity of the conventional perturbation approach breaks down at moderately high, but experimentally obtainable acoustic energy densities above $100~\SIJ/\SIm^3$ in combination with moderate thermal gradients above $1~\SIK/\SImm$. This need for an extension of the theory beyond perturbation theory is addressed in this Letter and in the accompanying detailed presentation of the nonperturbative model in Ref.~\cite{Joergensen2022}.

We introduce a nonperturbative iteration approach to investigate theoretically and numerically, the nonlinear effects appearing in a conventional acoustofluidic channel at high acoustic energy density $\Eac$, and we validate experimentally the model predictions. We take as our generic acoustofluidic model system, the widely used rectangular channel driven at resonance with a transverse half-wave standing acoustic wave, for which the streaming at low $\Eac$ is dominated by conventional boundary-driven streaming with four streaming rolls \cite{Barnkob2012a, Muller2013, Muller2014, Bach2018}. We show how nonlinear effects in the form of heating by viscous dissipation from the acoustic field inside the boundary layers, set up a steady temperature gradient. This gradient drives a strong thermoacoustic streaming in the bulk, which changes the streaming flow qualitatively from four to two flow rolls, and which by thermal convection alters the temperature field. Our analysis of this nonlinear phenomenon and its underlying mechanism fills a knowledge gap in nonlinear acoustics, and it provides a guidance for understanding and optimizing acoustofluidic systems running at high $\Eac$ such as high-intensity ultrasound focusing \cite{Solovchuk2013, Solovchuk2017, Sedaghatkish2020}, acoustic streaming-based micromixers \cite{Ozcelik2014, Patel2014, Bachman2020, Zhang2021}, particle manipulation devices \cite{Wiklund2012, Collins2016b, Collins2017}, and high-throughput acoustophoresis devices~\cite{Adams2012, Antfolk2015, Urbansky2019}.

\begin{figure}[!t]
\centering
\includegraphics[width=1.0\columnwidth]{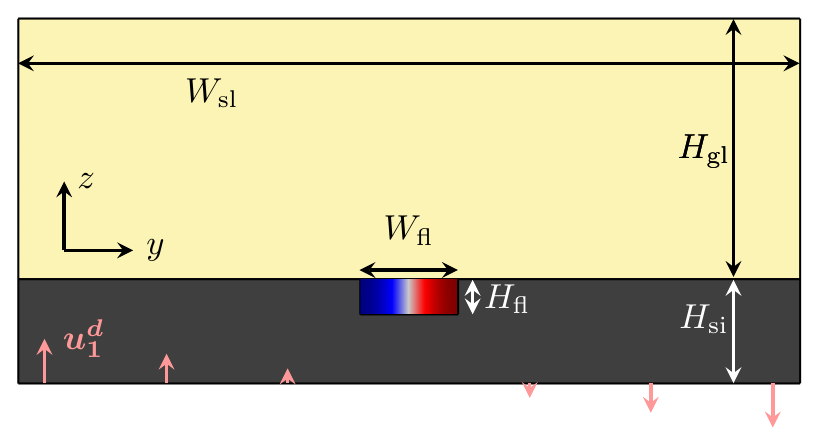}
\caption[]{\figlab{sketch}
Cross-section of the modeled rectangular channel of width $W_\fl=375~\SImum$ and height $H_\fl=135 \, \SImum$ embedded in a silicon base of width $W_\sl = 3~\SImm$ and height $H_\mathrm{si} = 0.4~\SImm$, and covered by a glass lid of height $H_\mathrm{gl} = 1~\SImm$. The bottom-edge actuation is the displacement $\uuu_1^d = d_0(y/W_\mr{sl})\:\eee_y$ with an adjustable amplitude~$d_0$ at frequency $f_0=1.911~\SIMHz$.
}
\end{figure}

\textit{Physical model.---}We consider a long straight microchannel with a cross-sectional of width $W_\textrm{fl}=375~\SImum$ and height $H_\textrm{fl}=135~\SImum$ embedded in a silicon base with a glass lid, see \figref{sketch} and Refs.~\cite{Joergensen2022, Barnkob2012a, Muller2013, Muller2014, Bach2018}. To excite the horizontal half-wave resonance mode in the fluid, the system is actuated at frequency $f_0=1.911~\SIMHz$ by the bottom-edge displacement sketched in \figref{sketch}. The response to the acoustic actuation is governed by the conservation equation for mass, momentum, and energy in the fluid and solid. The independent fields are the pressure $p$, the velocity $\vvv$, and the temperature $T$ in the fluid, and the displacement $\uuu$ and $T$ in the solid.

We study a fluid characterized by the following material parameters: density $\rho$, isothermal compressibility $\kapT$, thermal conductivity $\kth$, specific heat $\cp$, dynamic and bulk viscosity $\eta$ and $\etab$, thermal expansion coefficient $\alphap$, the ratio of specific heats $\gamma = \cp/\cv$, and the isentropic and isothermal compressibility $\kaps$ and $\kapT=\gamma \kaps$. The temperature dependence of the parameters for water are given by the polynomials derived in Ref.~\cite{Muller2014}. The elastic solid is characterized by density $\rho$, longitudinal and transverse sound speed $c_\mathrm{lo}$ and $c_\mathrm{tr}$, thermal conductivity $\kth$, thermal expansion coefficient $\alphapO$ and isothermal compressibility $\kapT$~\cite{Joergensen2022}.

We use the iterative thermoviscous model presented in our concurrent paper~\cite{Joergensen2022}, a model that is nonperturbative unlike the perturbative models traditionally used in acoustofluidics \cite{Muller2014, Joergensen2021}. The model exploits that the acoustic fields varies much faster ($\sim 10^{-7}~\SIs$) than the hydrodynamic and thermal flows ($\sim 10^{-2}~\SIs$), so that the fast and slow dynamics can be solved separately. Here, we study the stationary limit of the slow time scale and describe any given physical field $Q_\mr{phys}$ as a sum of a stationary field $Q_0$ and a time-varying acoustic field $\re\big\{Q_1 \ee^{-i\omega t}\big\}$ with a stationary complex-valued amplitude $Q_1$,
 \begin{equation} \eqlab{Qphys}
 Q_\mr{phys}(t)=Q_0+\re\big\{Q_1 \ee^{-i\omega t}\big\}.
 \end{equation}
A product of two acoustic fields will contain a steady part $\avr{a_1 b_1}=\frac12 \re \big\{a_1 b_1^* \big\}$ where the asterisk denotes complex conjugation. We neglect higher harmonics with angular frequency $n \omega$, $n=2,3,\ldots$. In Ref.~\cite{Joergensen2022}, we use this ansatz to separate the governing equations in a set that controls the acoustic fields, and a set that controls the stationary fields.

Acoustofluidic systems also exhibit dynamics on two different length scales, one set by the wavelength of the acoustic fields and one set by the viscous and thermal boundary layers. The boundary conditions on the velocity field, stress, heat flux, and temperature at the fluid-solid interface results in the appearance of thermal boundary layer of width $\delt$ in both the fluid and the solid, and in a viscous boundary layer of width $\dels$ in the fluid. These boundary layers are localized near fluid-solid interfaces, and their dynamically-defined widths (jointly called $\delta$) are small compared to a typical device size or wavelength $d$ \cite{Karlsen2015}, $\dels = \sqrt{2 \nu_0/\omega}$ and $\delt \approx \sqrt{2\Dth/\omega}$. Typically, $\delt \lesssim \dels \lesssim 500~\SInm$, which is more than two orders of magnitude smaller than $d\sim100~\SIum$. In our model \cite{Joergensen2022}, we use $\delta \ll d$ to separate the fields into a bulk field ($d$) and boundary layer field ($\delta$), that are connected through the boundary conditions at the fluid-solid interface.
The model presented in Ref.~\cite{Joergensen2022} uses the separation of time and length scales to setup an iterative model with effective boundary conditions that enables simulations of nonperturbative acoustofluidic systems without numerically resolving the viscous and thermal boundary layers. All dependent fields are given analytically by the independent fields as shown in Ref.~\cite{Joergensen2022}.

\textit{Acoustic fields.---}In the thermoviscous model of Ref.~\cite{Joergensen2022}, the acoustics is fully described by the pressure field $p_1$ in the fluid and by the displacement field $\uuu_1^d$ in the solid through the Helmholtz and Cauchy equations,
 \begin{subequations} \eqlab{acousticsEqu}
 \begin{align}
 \eqlab{fast_fluid}
 \nabla^2 p_1 & = -\kcsqr p_1,
 \quad \kc = \frac{\omega}{c_0}(1+\ii\GamOcfl),
 \\
 \eqlab{fast_solid}
 -\rho  \omega^2 \uuu_1^d &= \nablabf \cdot \sigmabf_1^{d,\mr{sl}},
 \\
 \sigmabf_1^{d,\mr{sl}} &= -\frac{\alfP}{\kapT} T_1 \III +  \rho\cTsqr \Big[\nablabf\uuu_1 + (\nablabf\uuu_1)^T\Big]\nn\\
 &\quad + \rho\big(\cLsqr-2\cTsqr\big)(\nablabf \cdot \uuu_1) \III,
 \end{align}
 \end{subequations}
where $\sigmabf_1$ is the stress tensor. $p_1$ and $\uuu_1^d$ are connected through effective boundary conditions on the fluid-solid interface taking the boundary layers into account analytically: a no-slip condition for the velocity and a continuity condition on the stress, as described in Ref.~\cite{Joergensen2022}. The acoustic velocity $\vvv_1$ and the temperature field $T_1$ are given by $p_1$ and the analytical boundary-layer fields \cite{Joergensen2021}.

\textit{Stationary fields.---}The stationary fields are the pressure $p_0$ and streaming velocity $\vvv_0^d$ in the fluid, and the temperature $T_0^d$ in the solid and fluid. The acoustic timescale affects $T_0^d$, $p_0$, and $\vvv_0^d$ through bulk terms (heating and the acoustic body force $\fffac^d$) and corrections due to the boundary layers to the boundary conditions. The governing equations for $\vvv_0^d$ and $p_0$ are \cite{Joergensen2022},
 \begin{subequations} \eqlab{stationaryEqu}
 \begin{align}
 \eqlab{v0d_gov}
 0&=\nablabf \cdot  \vvv_0^d ,\\
 0&=- \nablabf\Big[p_0^d -\avr{\Lac^d} \Big] + \eta_0 \nabla^2 \vvv_0^d -\nablabf\cdot\Big[\rho_0 \vvv_0 \vvv_0\Big] + \fffac^d.
 \\
 \eqlab{facd}
 \fffac^d &= -\frac{1}{4}\abs{\vvv_1^{d,p}}^2\nablabf \rho_0 -\frac{1}{4}\abs{p_1}^2\nablabf\kaps \nn\\
 &\quad +\left(1-\frac{2 a_\eta (\gamma-1)}{\beta+1} \right)\frac{\Gamma \omega}{c_0^2} \avr{\vvv_1^{d,p} p_1}
 \nn\\
 &\quad +2a_\eta \eta_0 (\gamma -1) \frac{\omega}{c_0^2} \avr{\ii\vvv_1^{d,p} \cdot \nablabf \vvv_1^{d,p}}.
\end{align} \end{subequations}
The boundary layers generate a slip velocity on the solid-fluid interface, and the resulting effective boundary condition on $\vvv_0^d$ is~\cite{Joergensen2022},
 \begin{subequations} \eqlab{v0d_AB}
 \begin{align}
 \vvv_0^{d0} &=  \left(\AAA \cdot \eee_x \right)\eee_x + \left(\AAA \cdot \eee_y \right)\eee_y+ \left(\BBB \cdot \eee_z \right)\eee_z,
 \end{align}
 \begin{align}
 \AAA &= - \frac{1}{2 \omega} \mr{Re}\Bigg{\{} \vvv_1^{\delta 0 *} \cdot \nablabf \left(\frac{1}{2} \vvv_1^{\delta 0} -\ii \VVV_1^0 \right) -\ii \VVV_1^{0*} \cdot \nablabf \vvv_1^{d,p} \nn
 \\
 &\quad+\left[ \frac{2-i}{2} \nablabf \cdot \vvv_1^{\delta 0 *} + i \left( \nablabf \cdot \VVV_1^{0 *} - \partial_z v_{1z}^{d *}\right)\right]\vvv_1^{\delta 0 }\Bigg{\}}\nn
 \\
 &\quad+ \frac{1}{2 \eta_0} \mr{Re}\Big{\{} \eta_1^{d0} \vvv_1^{\delta 0*} +\frac{1}{1-\ii \dels / \delt} \eta_1^{\delta 0}  \vvv_1^{\delta 0*}  \Big{\}},
 \end{align}
 \begin{align}
 \BBB &= \frac{1}{2 \omega}  \mr{Re}\Big\{ i \vvv_1^{d 0 *} \cdot \nablabf \vvv_1^{d,p} \Big\}.
 \end{align}
 \end{subequations}
Here, $\VVV_1^0$ is the velocity of the fluid-solid interface, $\eee_z$ is the surface normal vector, and $\eee_x$ and $\eee_y$ are perpendicular to $\eee_z$. The streaming flow $\vvv_0^d$ can be driven either by the acoustic body force $\fffac^d$, called bulk-driven streaming, or by the effective boundary condition on $\vvv_0^{d0}$, called boundary-driven streaming.

The stationary temperature $\TOdfl$ in the fluid is governed by the heat equation (energy conservation)~\cite{Joergensen2022},
 \begin{subequations} \eqlab{slow_T_fluid}
 \begin{align}
 0&= -\nablabf\cdot \Big[\kthO \nablabf T_0^{\fl,d}\Big]-\cp\rho_0\vvv_0 \cdot \nablabf T_0^{\fl,d}+\Pac^d,\\
 \Pac^d &= -\nablabf\cdot \Big[\avr{\kthId \nablabf T_1^d} -\avr{p_1 \vvv_1^{d,p}}+ \avr{\vvv_1^{d,p}\cdot \taubf_1^d}\Big]\nn\\
 &\quad
 - \cp \avr{\rho_1^d\vvv_1^{d,p}}\cdot \nablabf T_0^d,
 \end{align} 
 \end{subequations}
and similarly for $\TOdsl$ in the solid~\cite{Joergensen2022},
 \begin{subequations} \eqlab{slow_T_solid}
 \begin{align}
 0&= - \nablabf \cdot \left[ \kthO \nablabf T_0^{\sl,d} \right] +\Pac^d\\
 \Pac^d &= -\nablabf\cdot\avr{\kthIsl \nablabf T_1^{sl,d}}
 \end{align} 
 \end{subequations}
Here, $\Pac^d$ is the power density delivered by the acoustic wave through frictional dissipation and energy flux. $\TOdfl$ and  $\TOdsl$ are connected at the fluid-solid interface by the two effective boundary conditions taking the boundary layers into account analytically: continuity of temperature and of heat flux, applied respectively as a Dirichlet condition on $\TOdfl$ and a flux condition on $\nnn\cdot\nablabf\TOdsl$ \cite{Joergensen2022},
 \begin{subequations} \eqlab{T0BC}
 \begin{align}
 \eqlab{bc_T0}
 &T_0^{\fl ,d} = T_0^{\sl ,d} - T_0^{\fl,\delta 0} \nn\\
 & \quad
 - \frac{1}{2}\re\Big\{ \uuu_1 \cdot \nablabf T_1^{\fl,d*}-\kt^\fl (\uuu_1\cdot \nnn) T_1^{\fl,\delta0*}\Big\},
 \\
 \eqlab{bc_heat}
 &\kthOsl \nnn \cdot \nablabf T_0^{\sl,d} = \kthOfl \nnn \cdot \nablabf T_0^{\fl,d}
 +\kthOfl \partial_z T_0^{\fl,\delta} \nn\\
 & \quad
 +\frac{1}{2}\re\Big\{\kt^\fl \kthIfl T_1^{\fl,\delta*}
 -\frac{2  \ii }{\delt^2} \kthOfl(\uuu_1 \cdot \nnn)   T_1^{\fl,\delta*} \Big\}
 \end{align} 
 \end{subequations}
In summary, the bulk temperature $T_0^d$ is governed by the heat equations~\eqsnoref{slow_T_fluid}{slow_T_solid} together with the effective boundary conditions~\eqnoref{T0BC}.

\begin{figure*}[!t]
\centering
\includegraphics[width=1.0\textwidth]{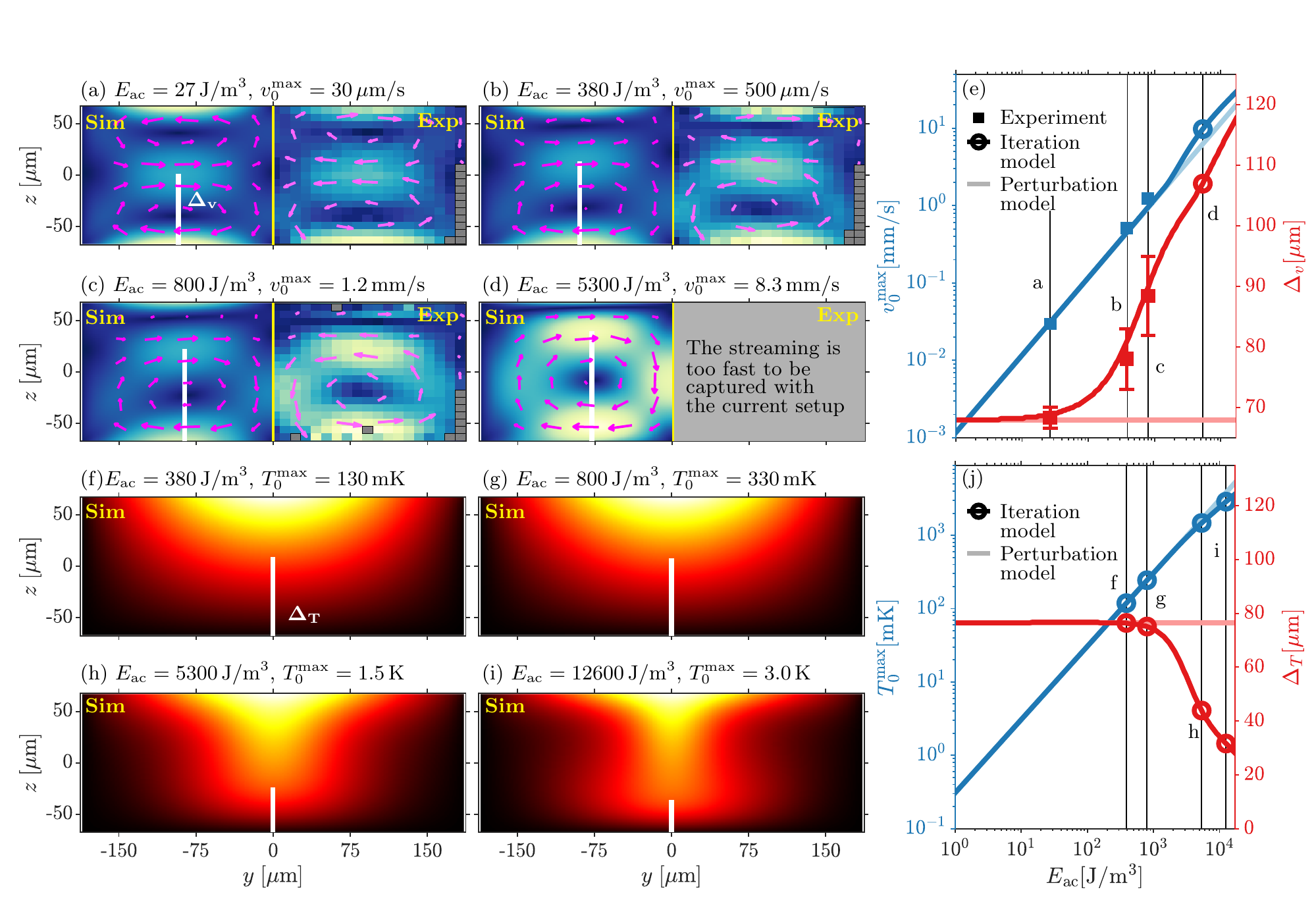}
\caption[]{\figlab{results}
Simulation and experimental results for $\vvv_0$ and $T_0$ in the microchannel. (a)-(d) Vector of $\vvv_0$ and color plot of $|\vvv_0|$ from 0 (blue) to $v_0^\mr{max}$ (yellow) at four  $\Eac$. For each $\Eac$, the left half are simulations and the right half are experiments. $\Delta_v$ (white bar) is the height where $v_{0y}$ is maximal. (e) Plots of  $\vvv_0^\mr{max}$ and $\Delta_v$  vs.\ $\Eac$   with data from simulations and experiments, showing the transition from boundary- to bulk-driven streaming. The error bars on experimental $v_0^\mr{max}$ and $\Eac$ are within the square markers. The round markers represent the simulations shown in panels (d) and (f)-(i). (f)-(i) Color plot of simulated temperature increase $T_0$ from 0 (black) to $T_0^\mr{max}$ at four $\Eac$. $\Delta_T$ (white bar) is the height where $\TO = \frac12 T_0^\mr{max}$. (j) Plots of simulated $T_0^\mr{max}$ and $\Delta_T$ vs.\ $\Eac$ showing a transition from diffusion- to convection-dominated heat transport.
}
\end{figure*}

\textit{Experimental method.---}The experiments were performed using a long straight microchannel of width $W = 375~\SImum$ and height $H = 135~\SImum$ in a glass-silicon chip with a piezoelectric transducer glued underneath. The transducer was driven at a frequency of 1.97~MHz at input power $P_\mathrm{in}$ = 6.1, 86.8, and 182.5~mW, resulting in the energy density $\Eac$ = $27.2\pm1.1$, $388.7\pm15.9$, and $817.3\pm33.5~\SIJ/\SIm^3$, respectively, as measured from the focusing of 5.0-$\SImum$-diameter particles at 140~fps using confocal micro-particle image velocimetry (\textmu PIV) at the low $P_\mr{in}$ \cite{Qiu2019}. At higher $P_\mr{in}$, $\Eac$ is estimated using the proportionality $\Eac \propto P_\mr{in}$. The confocal \textmu PIV technique only captures the particle motion near the focal plane (channel mid-height), excluding particles near the top and bottom walls influenced by hydrodynamic and acoustic particle-wall interactions, and as a result, $\Eac$ is measured accurately. The acoustic streaming for each $\Eac$ was measured at 10 to 60~fps by tracking the motion of 0.5-$\SImum$-diameter particles using a defocusing-based 3D particle tracking technique~\cite{Barnkob2015, Barnkob2020, Rossi2020}. To avoid the resonance frequency shift due to the temperature rise of the transducer under moderate (86.8~mW) and high (182.5~mW) $P_\mathrm{in}$, each measurement was run for 2~s and repeated 100 times to improve the statistics, resulting in 7800-12000 recorded frames for each driving condition.

\textit{Results and discussion.---}The simulation and experimental results shown in \figref{results} reveal the dominant nonlinear behavior of the stationary streaming $\vvv_0^d$ and temperature $T_0$ in a standard acoustofluidic device. In the linear regime at low $\Eac \lesssim 30~\SIJ/\SIm^3$, $\vvv_0^d$ is dominated by the boundary-driven streaming entering the model through the slip-velocity condition~\eqnoref{v0d_AB}, and the usual four boundary-driven streaming rolls appear, see \figref{results}(a). Due to friction in the viscous boundary layers, heat is generated both at the top and bottom of the channel. At the bottom, this heat is removed efficiently because of the high heat conductivity of silicon. At the top, however, the heat is removed less efficiently by the lower heat conductivity of glass, and a steady temperature gradient $\nablabf T_0$ is established, which explains the temperature $T_0^d$ seen in \figref{results}(f).

The gradient $\nablabf T_0$ created by the acoustic frictional heating results in gradients in $\nablabf \rho_0$ and $\nablabf \kappa_0$, thus inducing a thermoacoustic body force \eqnoref{facd} $\fffac^d$ \cite{Joergensen2021, Qiu2021},
 \begin{align}
 \fffac^d & \approx
 -\frac{1}{4}\vert p_1\vert^2 \nablabf \kappa_{s,0}- \frac{1}{4} \vert \vvv_1\vert^2 \nablabf \rho_0
 \eqlab{fac}
 \\ \nn
 & =
 -\frac{1}{4}\vert p_1\vert^2 \Big(\frac{\partial \kappa_s}{\partial T}\Big)_{T_0} \nablabf T_0
 - \frac{1}{4} \vert \vvv_1\vert^2  \Big(\frac{\partial \rho}{\partial T}\Big)_{T_0}\nablabf T_0.
 \end{align}
Since $|p_1|^2 \propto |\vvv_1|^2 \propto \Eac$ and $|\nablabf T_0| \propto \Pac^d \propto \Eac$, we have $|\fffac^d| \propto \Eac^2$, and $\fffac^d$ will become important in the bulk at high $\Eac$ and cause qualitative nonlinear changes of the streaming pattern. According to \eqref{fac}, $\fffac^d$ is pointing toward high temperature at the top and is strongest at the pressure antinodes at the sides \cite{Joergensen2021, Qiu2021}. Consequently, $\fffac^d$ pushes liquid from the sides up toward the top center, with a back-flow down along vertical center axis, thus creating a streaming pattern that consists of two streaming rolls in each side of the channel. This  pattern is seen at the high $\Eac=5300~\SIJ/\SIm^3$ in \figref{results}(d), where the streaming is completely dominated by the thermoacoustic streaming. The transition from boundary-driven streaming at low $\Eac$ to bulk-driven streaming at high $\Eac$ is studied qualitatively in \figref{results}(a)-(d) and quantitatively in \figref{results}(e). During the transition, the two bottom streaming rolls expand and the two top rolls shrink, see \figref{results}(b)-(c) at $\Eac=380$ and $800~\SIJ/\SIm^3$, respectively. The bottom rolls expand because they rotate the same way as the two thermoacoustic streaming rolls.

This transition is studied quantitatively in \figref{results}(e) by plotting the maximum streaming velocity $v_0^\mr{max}$ and the vertical distance $\Delta_v$ (thick white line) from the bottom of the channel to the position of the maximum horizontal streaming velocity $\mr{max}(v_{0y})$ toward the center occurs. In the log-log plot (dark blue), the perturbative result $v_0^\mr{max} \propto \Eac$ is valid up to $\Eac \approx 1000~\SIJ/\SIm^3$, but at higher values $v_0^\mr{max}$ increases faster. A stronger signal is seen in the log-lin plot (dark red), where the perturbative result $\Delta_v \propto \Eac^0$ only holds for $\Eac \lesssim 30~\SIJ/\SIm^3$, after which point $\Delta_v$ increases with increasing $\Eac$.

As the streaming velocity increases, convection becomes increasingly important for the heat transport~\eqnoref{slow_T_fluid} and strongly affects the temperature field, see \figref{results}(f-i) for $\Eac = 380$, 800, 5300, and $12,600~\SIJ/\SIm^3$. Convection becomes important at a P\'eclet number $\mathrm{Pe}= \abs{\vvv_0} H_\textrm{fl} / \Dth\sim 1 $ corresponding to $\abs{\vvv_0}\sim 1~\SImm/\SIs$, consistent with \figref{results}(f--j). Qualitatively, we see that for $\Eac \gtrsim 800~\SIJ/\SIm^3$, the two flow rolls pull the temperature profile down along the vertical center axis. We quantify this effect by the maximum temperature $T_0^\mr{max}$ and the vertical distance $\Delta_T$ along the center axis from the bottom edge to the point where $T_0 = \frac12 T_0^\mr{max}$. The thermoacoustic streaming increases the heat transport from the fluid-glass interface to the silicon wafer, thus $T_0^\mr{max}$ increases less steeply than the perturbative result, $T_0^\mr{max} \propto \Eac$, as seen in the log-log plot (blue) of $T_0^\mr{max}$ vs.\ $\Eac$ for $\Eac \gtrsim 5000~\SIJ/\SIm^3$ in \figref{results}(j). A stronger signal is seen in the log-lin plot (dark red), where the perturbative result $\Delta_T \propto \Eac^0$ only holds for $\Eac \lesssim 500~\SIJ/\SIm^3$, after which point $\Delta_T$ decreases with increasing $\Eac$.

\textit{Conclusion.---}In this Letter we have shown numerically and experimentally that the acoustic streaming in a standard microscale acoustofluidic device is changed qualitatively for moderately high acoustic energy densities $\Eac \gtrsim 500~\SIJ/\SIm^3$. We have explained this effect by a nonperturbative model \cite{Joergensen2022}, in which a transition from boundary- to bulk-driven acoustic streaming occurs, as the acoustic body force $\fffac$ begins to dominate the streaming at increased $\Eac$ due to the internal heating generated in the viscous boundary layers. We have shown good qualitative and quantitative agreement between our model predictions and experimental data.

$\Eac \gtrsim 500~\SIJ/\SIm^3$ can easily be obtained in standard acoustofluidic devices, where $\Eac \approx 10 - 50~\SIJ/\SIm^3 \times \big[U_\mr{pp}/(1~\SIV)\big]^2$ has been reported in the literature, $U_\mr{pp}$ being the applied voltage on the piezoelectric transducer \cite{Barnkob2010, Augustsson2011, Barnkob2012, Muller2013}. The physical understanding of how such acoustofluidic devices behave at high $\Eac$ is important for the continued development of high throughput devices in particular for biotech applications.

\begin{acknowledgments}
We thank R. Barnkob and M. Rossi for providing the software DefocusTracker, \url{defocustracking.com/}. WQ was supported by MSCA EF Seal of Excellence IF-2018 from Vinnova, Sweden's Innovation Agency (Grant No. 2019-04856). HB and JHJ was supported by Independent Research Fund Denmark, Natural Sciences (Grant No. 8021-00310B).
\end{acknowledgments}

%
%

%

\end{document}